\documentclass[lettersize,journal,compsoc]{IEEEtran}
\usepackage{amsmath,amsfonts}
\usepackage{algorithmic}
\usepackage{algorithm}
\usepackage{array}
\usepackage{url}
\usepackage[caption=false,font=normalsize,labelfont=sf,textfont=sf]{subfig}
\usepackage{textcomp}
\usepackage{stfloats}
\usepackage{url}
\usepackage{verbatim}
\usepackage{graphicx}
\usepackage{balance}
\usepackage{graphicx} 
\usepackage{url}
\usepackage{multirow, makecell}
\usepackage[breakable,skins]{tcolorbox}
\usepackage{xcolor}
\usepackage{enumitem}
\usepackage{tabularx}
\usepackage{hyperref}
\usepackage{xspace}
\usepackage{mathpazo}
\newcommand{\palatino}{\fontfamily{ppl}\selectfont}

\newcommand{\thickhline}{
    \noalign{\hrule height 2pt}
}

\newcommand{\TCD} {Toxic Conversations Dataset}
\newcommand{\DTD} {Derailed Toxic Dataset}
\newcommand{\NTD} {Non-Toxic Conversations Dataset}

\newcommand{\EP} {external participants\xspace}
\newcommand{\PC} {project contributors\xspace}

\definecolor{colortitlebg}{HTML}{004D40}
\definecolor{colortitletext}{HTML}{FFFFFF}
\definecolor{colorcommentbg}{HTML}{C8E6C9}
\definecolor{colorcommenttext}{HTML}{000000}

\newcounter{observationcomment@counter}
\newenvironment{observecomment}[1][]{\refstepcounter{observationcomment@counter}
    \begin{tcolorbox}[adjusted title={Observation \#\arabic{observationcomment@counter}}, 
    fonttitle={\palatino\bfseries}, 
    enhanced jigsaw, 
    colbacktitle={colortitlebg},
    coltitle={colortitletext},
    arc=2pt,
    opacityframe=0,
    boxrule=0em,
    colback={colorcommentbg},#1,
    breakable]
    \color{colorcommenttext}
}{\end{tcolorbox}}

\newenvironment{promptbox}[2][]{
    \begin{tcolorbox}[title={#2},
    fonttitle={\palatino\bfseries}, 
    enhanced jigsaw, 
    colbacktitle=black,
    arc=2pt,
    opacityframe=0,
    boxrule=0.4mm,
    opacityframe=1, 
    colback=white,#1,
    breakable]
}{\end{tcolorbox}}

\title{Understanding and Predicting Derailment in Toxic Conversations on GitHub}

\author{
\IEEEauthorblockN{
Mia Mohammad Imran\IEEEauthorrefmark{1}, 
Robert Zita\IEEEauthorrefmark{2}, 
Rebekah Copeland\IEEEauthorrefmark{3}, 
Preetha Chatterjee\IEEEauthorrefmark{4}, 
Rahat Rizvi Rahman\IEEEauthorrefmark{5}, 
Kostadin Damevski\IEEEauthorrefmark{5}
}

\IEEEauthorblockA{\IEEEauthorrefmark{1}Missouri University of Science and Technology, Rolla, MO, USA\\
Email: imranm@mst.edu}

\IEEEauthorblockA{\IEEEauthorrefmark{2}Elmhurst University, Elmhurst, IL, USA\\
Email: rzita8729@365.elmhurst.edu}

\IEEEauthorblockA{\IEEEauthorrefmark{3}Eastern Mennonite University, Harrisonburg, VA, USA\\
Email: rebekah.copeland@emu.edu}

\IEEEauthorblockA{\IEEEauthorrefmark{4}Drexel University, Philadelphia, PA, USA\\
Email: preetha.chatterjee@drexel.edu}

\IEEEauthorblockA{\IEEEauthorrefmark{5}Virginia Commonwealth University, Richmond, VA, USA\\
Email: \{rahmanr12,kdamevski\}@vcu.edu}
}

\begin{document}

\maketitle

\begin{abstract}

Software projects thrive on the involvement and contributions of individuals from different backgrounds. However, toxic language and negative interactions can hinder the participation and retention of contributors and alienate newcomers. Proactive moderation strategies aim to prevent toxicity from occurring by addressing conversations that have derailed from their intended purpose. This study aims to understand and predict conversational derailment leading to toxicity on GitHub.

To facilitate this research, we curate a novel dataset comprising 202 toxic conversations from GitHub with annotated derailment points, along with 696 non-toxic conversations as a baseline. Based on this dataset, we identify unique characteristics of toxic conversations and derailment points, including linguistic markers such as second-person pronouns, negation terms, and tones of Bitter Frustration and Impatience, as well as patterns in conversational dynamics between project contributors and external participants.



Leveraging these empirical observations, we propose a proactive moderation approach to automatically detect and address potentially harmful conversations before escalation. By utilizing modern LLMs, we develop a conversation trajectory summary technique that captures the evolution of discussions and identifies early signs of derailment. Our experiments demonstrate that LLM prompts tailored to provide summaries of GitHub conversations achieve 70\% F1-Score in predicting conversational derailment, strongly improving over a set of baseline approaches.
\end{abstract}

\maketitle

\section{Introduction}

Toxicity is bad for the health of online communities, including those centered around software projects. Research demonstrates that toxic language significantly impedes the onboarding of newcomers into Open Source Software (OSS) projects~\cite{qiu2019signals, raman2020stress}. A 2017 GitHub survey revealed that 50\% of developers encountered negative interactions, with 21\% reporting that such experiences caused them to cease contributing~\cite{GitHubOpenSourceSurvey2017}. Consequently, it is imperative for projects to safeguard and promote the engagement of all participants, both newcomers and experienced contributors. 

Despite the increasing recognition of the negative impact of toxic interactions, existing toxicity detection methods are predominantly post-hoc~\cite{sarker2020benchmark, sarker2023automated, mishra2024exploring, raman2020stress}. These approaches identify and address toxic comments and behaviors only after they have occurred, often relying on manual moderation or automated tools that flag inappropriate content retrospectively. While post-hoc detection can mitigate some of the damage caused by toxic interactions, it fails to prevent the initial harm and allows negative behaviors to persist unchecked for extended periods. This reactive approach not only delays intervention but also burdens community moderators and risks alienating contributors who might have otherwise remained engaged. Consequently, there is a pressing need for proactive solutions that can anticipate and preemptively address potential toxicity.

Proactive moderation in OSS projects involves moderators engaging in discussions to encourage positive behavior among developers. This approach contrasts with reactive moderation strategies, which include removing comments or locking threads post-incident. Effective proactive moderation can preempt toxicity; however, it is impractical for moderators to continuously monitor the multitude of communication channels (e.g., issues, chats, discussion boards) within an OSS project~\cite{schluger2022proactive}. On the other hand, automatic proactive moderation necessitates a profound understanding of the specific context and community dynamics. Unlike platforms such as X (formerly Twitter) or Reddit, GitHub projects often exhibit more subtle inappropriate behaviors, such as entitlement, miscommunication, or resistance to new practices, rather than overt aggression~\cite{hsieh2023nip}.

The advent of foundational LLMs (e.g., GPT, LLaMa), capable of comprehending human text, offers a unique opportunity to integrate advanced NLP techniques into OSS environments to proactively detect and mitigate potential communication issues. 

This paper aims to understand the characteristics of toxic conversations on GitHub and how conversations derail into toxicity. We use this knowledge to provide a method for automatically detecting if conversations will derail. More specifically, the paper makes the following key contributions:

\begin{itemize}
    \item We curate a dataset of 202 toxic conversations on GitHub with annotated derailment points, as well as 696 non-toxic conversations as a baseline.

    \item We examine the characteristics of toxic conversations and derailment points, identifying specific unique characteristics of each.

    \item We present an automated approach to OSS moderation that predicts whether a conversation will derail into toxicity. It is based on a novel prompt methodology that generates Summaries of Conversation Dynamics (SCD) for GitHub conversations. Our approach is able to effectively detect early signs of conversational derailment, achieving an F1-score of 0.70.
\end{itemize}

Our study's datasets, scripts, and output logs are publicly available online at URL: \color{blue}\url{https://anonymous.4open.science/r/derailment-oss-replication-C8B1}\color{black}.

\section{Example of Conversational Derailment on GitHub}

\begin{figure}[t]
\centering
\includegraphics[width=0.98\linewidth]{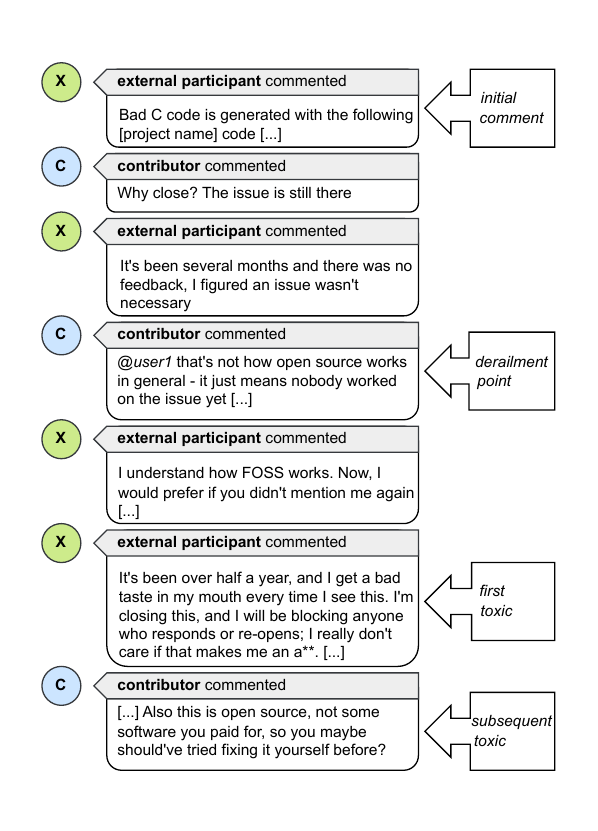}
\caption{Example of a toxic conversation on GitHub.}
\label{fig:motivating_example}
\end{figure}

When a conversation on GitHub channels, like issues or pull requests, turns toxic, the toxicity often occurs with identifiable signs in the previous comments. In this research, we focus on understanding the early signs that a conversation will turn toxic on GitHub. These preceding comments, where it becomes clear that the conversation has moved away from being productive and taken a turn towards negativity, are called {\em derailment points}~\cite{zhang2018conversations}. 

Overall, toxic conversations often contain the following identifiable elements: 1) a conversation-initiating comment, 2) a derailment point comment, 3) a first toxic comment, and 4) (zero or more) subsequent toxic (or non-toxic) comments. Figure~\ref{fig:motivating_example} shows an example of a toxic conversation, highlighting these different structures. In this conversation between an OSS project contributor and an external participant (i.e., someone who has never made a commit to the repository), the contributor derails the conversation by making a mocking comment. The external participant responds with frustration and then makes a toxic, insulting remark. This is followed by another toxic comment, this time made by the contributor.

\section{Characteristics of Toxic Conversations on GitHub}

Understanding the characteristics of toxic conversations is crucial for developing effective intervention strategies. This section aims to identify and analyze the key properties of toxic conversations.
We identify these properties by examining various aspects such as participants' roles, comment patterns, thread initiation, linguistic features, and comment timing. Our analysis focuses on understanding the conversational patterns that lead to toxicity and the behaviors associated with it. 

\subsection{Datasets}

We curated two datasets of conversations, one toxic and one non-toxic, sourced from GitHub issues and pull requests. We describe our annotation process for each dataset, where our primary goal was to include high-quality annotated and representative examples of GitHub conversations.












\begin{table}
\caption{Definitions and examples of uncivil tone-bearing discussion features (TBDF).}
\centering
\footnotesize
    
\begin{tabular}{|>{\raggedright\arraybackslash}m{1.3cm}|>{\raggedright\arraybackslash}m{3.12cm}|>{\raggedright\arraybackslash}m{3.12cm}|}
\hline
\textbf{TBDF} & \textbf{Definition} & \textbf{Example}\\ \thickhline

Bitter\newline Frustration
& Expressing strong frustration, displeasure, or annoyance	
& \textit{No answer, no reaction, what kind of support is that.}\\ \hline

Impatience
& Expressing dissatisfaction due to delays	
& \textit{Issue not fixed in 30 days? Must be gone!}\\ \hline

Mocking
& Ridiculing or making fun of someone in a disrespectful way 
& \textit{Legend says this issue will still exist even on the end of mankind.}\\ \hline

Irony
& Using language to imply a meaning that is opposite to the literal meaning, often sarcastically 
& \textit{Maybe you should actually write that down somewhere. You know, like in the documentation.} \\ \hline

Vulgarity
& Using offensive or inappropriate language 
& \textit{Who cares, same sh*t.} \\ \hline

Threat
& Issuing a warning that implies a negative consequence
& \textit{Any further responses will result in you being blocked from the repo entirely.} \\ \hline

Entitlement
& Expecting special treatment or privileges 
& \textit{[...] that's how good we are. I don't want your contribution. [...]} \\ \hline

Insulting
& Making derogatory remarks towards another person or project
& \textit{This looks like it was done by a 5 year old.} \\ \hline

Identity attacks/ Name-calling
& Making derogatory comments based on race, religion, gender, sexual orientation, or nationality
& \textit{I would not be surprised if this database is maintained by the [nationality].} \\ \hline
\end{tabular}%
\label{tab:uncivil_features}
\end{table}

\subsubsection{{\TCD}} 
We leverage a dataset recently released by Ehsani et al.~\cite{ehsani2024incivility}, which focuses on incivility in GitHub conversations. This dataset is based on 404 locked conversations (issues and PRs) on GitHub where the reason they were locked is listed as `too heated,' `spam,' or `off-topic.' These 404 conversation threads contain 5961 comments annotated with various categories of uncivil TBDFs (Tone Bearing Discussion Features). The definitions and examples of the incivility-related TBDFs are shown in Table~\ref{tab:uncivil_features}.

We use a LLM-aided model-in-the-loop annotation approach to identify the uncivil comments that are also toxic~\cite{bartolo2020beat}. 
Recent research shows that such a model-in-the-loop annotation methodology works well for this type of data, including hate and violent speech detection tasks~\cite{sanyal2024machines, zendel2024enhancing, gilardi2023chatgpt, nguyen2024human, wang2024human, zhu2024apt, jahan2024finding}.
For each uncivil comment (as annotated by Ehsani et al.), we asked GPT-4o if each comment is toxic or not. We provided the complete conversation until the current comment for context and the following definition to the LLM, \textit{``Toxicity is defined as 'rude, disrespectful, or unreasonable language that is likely to make someone leave a discussion'"}. GPT-4o identified 832 toxic comments belonging to 273 threads. Following GPT-4o's annotation, two of the paper's authors of the paper manually checked each toxic comment to ensure whether they were actually toxic or not. The two human annotators reached an initial agreement of 0.78 (Cohen's Kappa). The comments where the annotators disagreed were resolved through an in-person discussion.

After this annotation, we further excluded conversations where the first comment in the conversation was annotated as toxic. These are cases that are inappropriate for our analysis of conversational derailment in software engineering. Following this step, we retained 175 toxic GitHub threads. 

\subsubsection{\NTD} To analyze the properties of toxic conversations compared to ordinary, non-toxic GitHub issue/PR conversations, we collected a random sample of GitHub threads from the same repositories as the toxic threads. To reflect a realistic conversational data distribution, we intentionally sampled more data in this dataset by collecting four random threads for each toxic thread in the Toxic Dataset.

More specifically, we examined 15 threads before and 15 threads after each of the toxic threads. We included their data if they had at least two comments, were not marked as `too heated,' were not locked, or if locked, were marked as `resolved.' We then randomly selected four threads from the group. However, we were unable to collect all the data as some repositories had been removed since Ehsani et al.~\cite{ehsani2024incivility} collected their dataset, and some repositories did not have enough issues/PRs. Some of the conversations were also non-english, which we excluded. The resulting collection consisted of 723 GitHub issues and PR conversational threads.

To ensure these conversations are non-toxic, we used a similar model-in-the-loop approach as in curating the \TCD. Specifically, two authors of this paper manually reviewed the toxic comments identified by GPT-4o to determine whether each comment was truly toxic. They achieved a high agreement of 0.712 (Cohen’s Kappa), with the remaining disagreements resolved through in-person discussions. Ultimately, out of the initial 723 threads, 27 were identified as toxic, and the remaining 696 were marked as non-toxic, forming our \NTD.

We added the 27 toxic conversations identified through this process to the set of 175 \TCD, resulting in a total of 202 conversations containing 483 toxic comments. To ensure consistency with the existing conversations, two authors of this paper annotated TBDF for these 27 new toxic conversations following Ehsani et al.‘s annotation guidelines~\cite{ehsani2024incivility}. They initially achieved a Cohen’s Kappa of 0.70 and subsequently resolved their differences through discussion to reach complete agreement.

\subsection{Findings}

\subsubsection{Participants of Toxic Conversations}

\begin{figure}[t]
\centering
\includegraphics[width=0.99\linewidth]{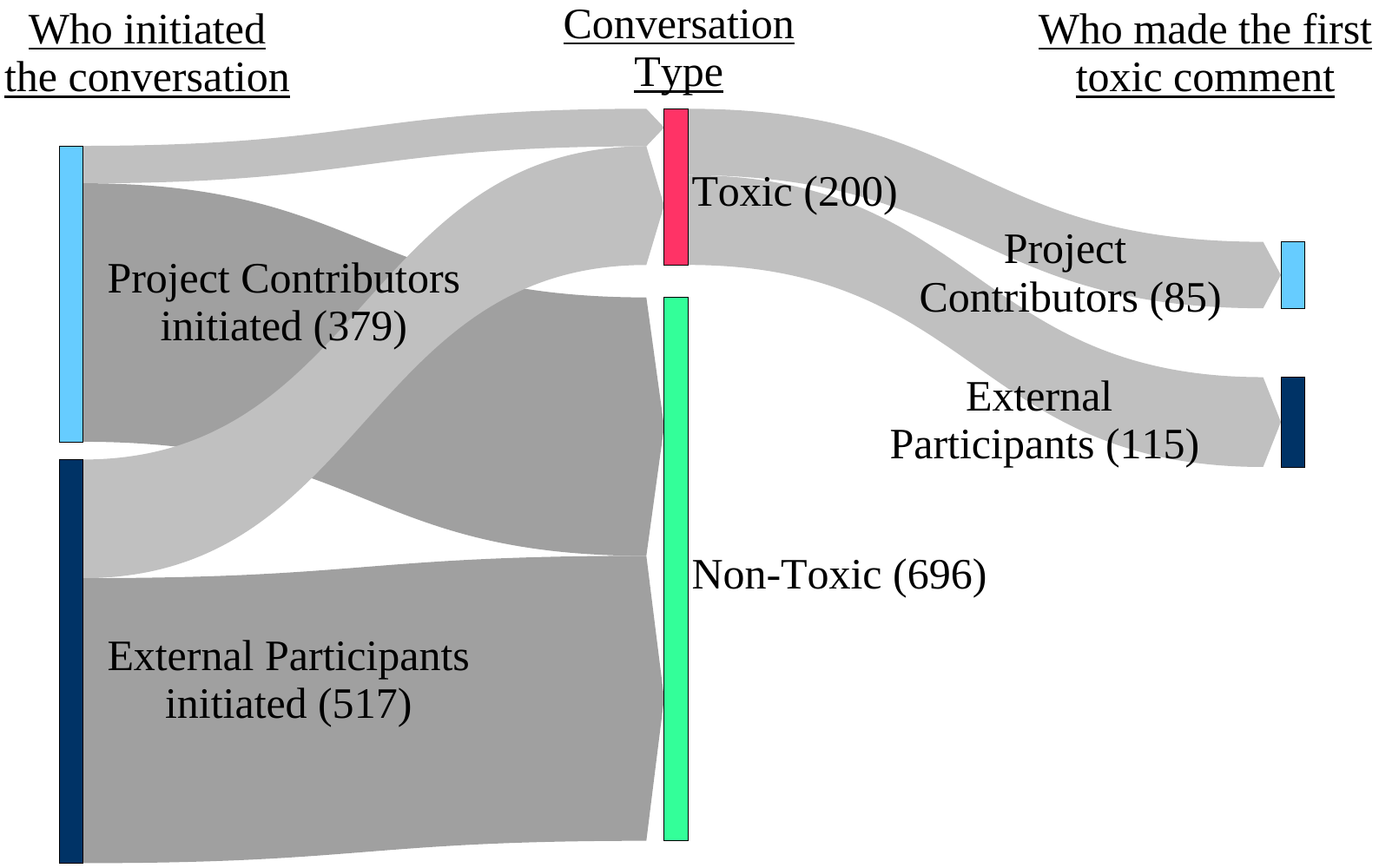}
\caption{Participants in different types of GitHub conversations.}
\label{fig:Toxic_conversations_initiated}
\end{figure}

We divide participants in the GitHub repositories into two categories: \textit{`Project Contributors'} and \textit{`External Participants'}. GitHub assigns specific roles such as Owner, Collaborator, Member, Contributor, or None~\footnote{\url{https://docs.github.com/en/graphql/reference/enums\#commentauthorassociation}}. Owners create the repository, Collaborators have administrative access, Members belong to the organization that owns the repository, and Contributors have made commits. The None role applies to authors without a specific association. GitHub offers a few additional roles that we did not encounter in either of our datasets. We classify the first four categories (Owner, Collaborator, Member, and Contributor) as \textit{Project Contributors} of the project, and take None to represent \textit{External Participants}. 
In the {\TCD}, there are 479 toxic comments made by 271 commenters in the None category, 96 in the Member category, 82 as Contributors, 22 as Collaborators, and 8 as Owners. Therefore, there are 271 {\EP} and 208 {\PC}. It is important to note, however, that these are not necessarily unique commenters. As multiple conversations were extracted from the same repositories, it is possible that individuals may appear in more than one conversation. We exclude two toxic conversations containing 4 toxic comments from this analysis because they were deleted from GitHub since Ehsani et al. published their dataset, and were therefore unavailable for the collection of author-related information. 

In these toxic conversations, we observe that there are slightly more comments made by {\EP} (52.79\% of comments) than by {\PC} (47.21\% of comments). However, in the non-toxic conversations, the percentage of {\PC} comments is higher (66.62\%). This suggests a slight shift in participation rates when toxicity is present, with \EP being more vocal in toxic conversations.

We observe that {\EP} initiate more (76.0\%, 152/200) of toxic conversations. This is much higher than the non-toxic conversations, where they initiated 52.44\% (365/696) of the conversations, as shown in Figure~\ref{fig:Toxic_conversations_initiated}. 
Despite {\PC} initiating only 24\% (48/200) of toxic conversations, they made a higher percentage of the first toxic comments (42.5\%, 85/200). This indicates that while developers are less likely to start toxic conversations, they are disproportionately responsible for the toxicity after the thread derails.

\begin{figure}[t]
\centering
\includegraphics[width=0.99\linewidth]{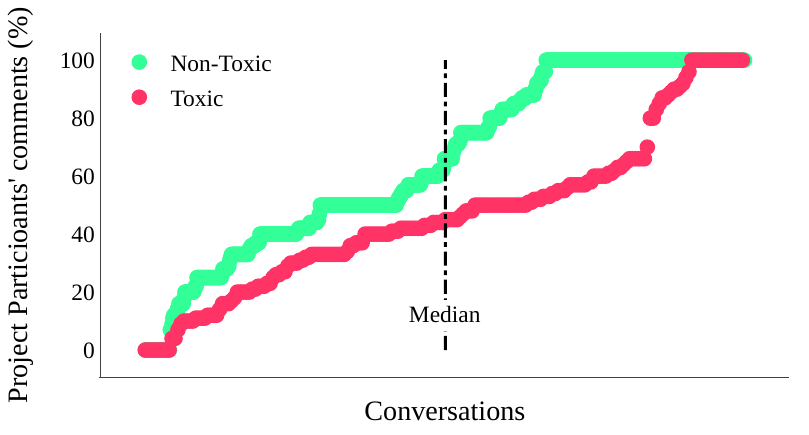}
\caption{Percentage of project participants' comments in GitHub conversation threads 
($N_{Toxic} = 202$; $N_{Non-Toxic} = 696$).}
\label{fig:develoeprs_comments}
\end{figure}

Figure~\ref{fig:develoeprs_comments} shows the {\PC} (vs. {\EP}) comment percentage in conversations from the {\TCD} vs {\NTD}. Each data point in the figure is an individual conversational thread. The median percentage of developer comments in 200 toxic conversations is 44.0\%, whereas for the 696 threads in the {\NTD} it is 66.0\%. 
Figure~\ref{fig:develoeprs_comments} also suggests that high developer engagement ($\geq$ 60\% of comments) in conversations appears to correlate with a reduction in thread toxicity. However, this relationship is not necessarily causal as other factors, such as the nature of the issue/PR, the tone set by the initial post, and the general community culture, may also influence the toxicity level of a thread.

\begin{observecomment}
External participants initiated 76.0\% of toxic conversations on GitHub and contributed 52.79\% of the comments in toxic threads, playing a significant role in driving toxicity. 
Conversations with higher {\PC} engagement tended to be less toxic.
\end{observecomment}

\subsubsection{Length of Toxic Conversations} Toxic conversations on GitHub tend to be relatively lengthy. The median number of comments in {\TCD} is 11, compared to the {\NTD} median of 6 comments. 

The median occurrence of the first toxic comment is 6 comments after the conversation begins. 
The distribution of the first toxic comment's occurrence is as follows: 24.75\% (50/202) of the time within the first 3 comments, 42.07\% (85/202) of the time within the first 5 comments, 56.93\% (115/202) of the time within the first 7 comments, and 70.79\% (143/202) of the time within the first 10 comments. The median number of comments after the first toxic comment is 3. Note that the moderators locked majority of these threads as heated, possibly preventing even more offensive comments from being posted.

We observe that 53.47\% (108/202) of the toxic conversations have more than one toxic comment with a median of 3 toxic comments per conversation. This indicates that toxic comments are likely to elicit more toxic replies.

\begin{observecomment}
Toxic conversations on GitHub tended to be longer, with a median of 11 comments versus 6 in the non-toxic conversations. Toxic comments usually appeared later on, at a median of 6 comments after the initial post. Over half of these conversations (53.47\%) had multiple toxic comments, indicating that toxicity often escalates.
\end{observecomment}

\subsubsection{Timing of Toxic Comments} 
The timing of toxic comments within a thread is important for understanding the dynamics of how toxicity emerges and escalates in conversations, as well as how it can be best mitigated~\cite{xia2020exploring}. We extracted the timestamp of each comment and calculated the time difference between consecutive comments. Table~\ref{tab:Toxic_comment_timing} illustrates the distribution of the timing for the first toxic comment across 202 threads in the {\TCD}. 
We made a number of observations:

\textit{Rapid Escalation}: The first toxic comment appears within an hour of the previous comment 51.98\% (105/202) of the time. Of these comments, 80.0\% (84/105) occur within a shorter timeframe than the median time between comments in each conversation thread, indicating that there is often a rapid response leading to toxicity.

\textit{Decreasing Likelihood Over Time}: The likelihood of a toxic comment occurring diminishes as more time passes after a comment is posted. For instance, only 9.90\% (20/202) of toxic comments appear within 1-3 hours and the proportion continues to decrease for longer intervals (see also Table~\ref{tab:Toxic_comment_timing}).

\textit{Delayed Toxicity Still Exists}: Despite the general trend of rapid escalation, a notable 9.41\% (19/202) of toxic comments occur after a week, highlighting that toxic interactions can resurface even after a period of inactivity.

\begin{observecomment}   
Toxic comments on GitHub often emerged quickly, with 51.98\% appearing within an hour of the previous comment and 80.0\% occurring earlier than the median response time in these threads. Despite the decreased likelihood of toxicity over time, 9.41\% of the first toxic comments appeared more than a week after the previous comment.
\end{observecomment}

\subsubsection{References to Other Participants in Toxic Comments}\label{toxic_addressing}

To identify patterns of how GitHub participants refer to each other in toxic comments, we examined two aspects: 1) mentioning someone using `@username,' and 2) quoting someone's previous comment.
Our analysis revealed that the first toxic comment in a conversation thread mentions someone using `@username' 25.74\% of the time, which is comparable to the average mentioning percentage of 25.0\% in the threads from the {\TCD}. However, a more notable difference emerges when examining the prevalence of quoting in toxic threads. The data indicates that 27.23\% of comments in toxic threads quote someone, which is substantially higher than the quoting rate of 11.72\% observed in the non-toxic threads. This finding suggests that quoting may serve as an indicator of contentious or confrontational interactions, potentially signifying a higher level of engagement with specific individuals or arguments.

The examination of common trigrams in toxic comments provides additional insights into the nature of these interactions. Phrases such as `you want to' and `if you want' frequently appear in toxic comments, indicating a tendency towards instructive, assumptive, or confrontational language directed at specific individuals.
Zhang et al. highlighted that the use of second person pronouns is strongly associated with nonconstructive disagreements~\cite{zhang2018conversations}. Our analysis of the first toxic comments on GitHub further supports this connection, revealing a substantially higher frequency of second person pronoun use (71.29\%, 144/202) in toxic comments than in all non-toxic threads' comments (35.99\%, 1913/5316). This stark contrast underscores the prevalence of direct address and potentially confrontational language in toxic exchanges. 
On the conversational level, we can observe similar properties as well: 202 toxic conversations have a median of 6 comments with second person pronouns while the 696 non-toxic conversations have a median of 2 comments. We observe that only 3.46\% (7/202) of toxic conversations did not have any second person pronoun use, compared to 11.93\% (83/696) in the {\NTD}.

The role of first person pronouns in non-constructive conversations, as noted by De Kock et al.~\cite{de2021beg} in the context of Wikipedia, is also supported by GitHub interactions. The first toxic comments exhibit slightly higher usage of first person pronouns (70.79\%, 143/202) compared to non-toxic threads' comments 60.68\%, 3226/5316). 

We observe that 55.94\% (113/202) of toxic comments use both first and second pronouns, compared to 24.66\% (1311/5316) of comments in the {\NTD}. This suggests that toxic exchanges may involve both direct address and self-referential language, potentially contributing to a more subjective and personal tone. 

\begin{table}[tb]
    \caption{Timeframe between Toxic comment and previous comment in the thread.}
    \centering
    
    \begin{tabular}{ | >{\raggedright}m{2.5cm} | l | >{\raggedright\arraybackslash}m{3cm} | }
        \hline
Passed time since previous comment & Count (\%) & Shorter than median timeframe in conversation\\ \thickhline
$<$ 1 hour & 105 (51.98\%) &  84/105 (80.0\%)\\ \hline
1-3 hours & 20 (9.90\%) & 10/20 (50.0\%)  \\ \hline
3-6 hours & 13 (6.44\%) & 5/13 (38.46\%)  \\ \hline
6-12 hours & 12 (5.94\%) & 4/12 (33.33\%)  \\ \hline
12-24 hours & 9 (4.55\%) & 1/9 (11.11\%) \\  \hline
1-7 days & 24 (11.88\%) & 6/24 (25.0\%)  \\ \hline
$>$ 1 week & 19 (9.41\%) & 0/19 (0\%)  \\ \hline
    \end{tabular}
    \label{tab:Toxic_comment_timing}
\end{table}

\begin{observecomment}
Toxic comments frequently quoted other commenters (27.23\% vs. 11.72\% in non-toxic threads) and used second person pronouns (71.29\% vs 35.99\%). 
Both first and second person pronouns occurred in 55.94\% of toxic comments compared to 24.66\% of regular comments.
\end{observecomment}

\subsubsection{Incivility TBDFs in Toxic Comments}\label{toxic_uncivil_tbdf}
Out of the 1025 comments containing incivility TBDFs in the 202 toxic conversation threads, a substantial 47.12\% (483/1025) are annotated as toxic. This high proportion demonstrates the close relationship between uncivil discourse features and toxicity, as observed by previous research~\cite{sadeque2019incivility}.
Notably, the TBDFs that exhibit a particularly strong association with toxicity are:
\begin{itemize}
\item Identity Attacks/Name-Calling: 84.62\% (22/26) toxic
\item Vulgarity: 78.33\% (47/60) toxic
\item Insulting: 77.44\% (127/164) toxic
\item Entitlement: 64.86\% (48/74) toxic
\end{itemize}
TBDFs that involve Identity Attacks, Vulgarity, or Entitlement are highly indicative of toxic interactions. Notably, Entitlement is especially prevalent in software engineering text, as shown by Miller et al.~\cite{miller2022did}. The prevalence of these TBDFs in toxic comments highlights the need to pay close attention to these specific features when identifying and addressing toxicity. 

On the other hand, conversely, some TBDFs show a lower correlation with toxic comments:
\begin{itemize}
\item Impatience: 18.35\% (29/158) toxic
\item Bitter Frustration: 31.17\% (115/368) toxic
\item Irony: 46.41\% (22/47) toxic
\end{itemize}
This suggests that expressions of frustration and impatience, although potentially detrimental to the conversation, may not always cross the threshold into overt toxicity. Mocking and Irony sometimes can be friendly banter or community jokes. 

In 75 cases where toxicity occurs suddenly, i.e., there is no derailment point, we observe that nearly half of the time (37/75) the reason for toxicity is the sudden occurrence of Vulgarity or Insulting TBDFs. This aligns with Miller et al.'s findings, where they noted over half of their sample of toxic comments (55\%) contained insulting, curse words or intentionally offensive language~\cite{miller2022did}.

\begin{observecomment}
We observed that 47.12\% of comments with uncivil TBDFs were toxic. TBDFs like Identity Attacks (84.62\%), Vulgarity (78.33\%), and Insults (77.44\%) strongly indicated toxicity, while Bitter Frustration (31.17\%) and Impatience (18.35\%) were less likely to be toxic.
\end{observecomment}

\subsubsection{Toxic Issue/Pull Request Labels} GitHub allows project maintainers to assign customized labels to conversations to provide a quick insight (e.g., `bug' or `needs triage'). Out of the 202 toxic issues/PRs, 107 (52.97\%) have labels assigned by participants. Using the GitHub API, we find that the largest label category is `bug' or similar (25.23\%, 27/107). This seems to indicate that conversations related to bugs are most prone to turn toxic. The second largest category is labels such as `feature', 'enhancement', and `suggestion' (17.76\%, 19/107). This is mostly because of disagreements over new ideas or changes that lead to heated discussions. Following this, `help wanted' appeared in 13 cases (12.15\%, 13/107), where one of the commenters assigned a label that indicated they needed assistance with some aspect of the project. Often, this type of conversation becomes toxic as a result of frustration from commenters. Labels such as `wontfix' and `rejected' occurred occasionally (12.15\%, 13/107), suggesting that hostility may be caused when a developer's PR or feature request is not accepted by the community. However, `positive status update' labels, such as `approved' and `completed' were still present in toxic issues (5.61\%, 6/107), demonstrating that although rejecting a feature request or PR often leads to toxicity, accepting it does not completely eliminate a negative response. 

\subsection{Implications}

This analysis of toxic conversations on GitHub has several implications. Firstly, it highlights that \EP are more likely to participate in and initiate toxic conversations than project contributors, in line with the findings of Miller et al.~\cite{miller2022did}. This difference in participation rates indicates a need for more focus on managing human interactions to curb toxicity. Additionally, this analysis shows that higher developer engagement correlates with lower toxicity levels, suggesting that developers' active participation can help maintain a constructive tone in conversations. Toxic conversations also tend to be longer, as seen in Xia et al.~\cite{xia2020exploring}, and toxic comments are often followed by more toxic comments~\cite{cheng2017anyone}, emphasizing the need for early detection and intervention to prevent further escalation. Communication patterns such as higher rates of quoting and the use of second-person pronouns~\cite{levy2022understanding}, are indicative of more direct and confrontational interactions, which can be flagged for early intervention. The rapid escalation of toxic comments, with responses often occurring within the hour, further highlights the need for real-time monitoring and quick responses. This occurrence of delayed toxicity suggests that unresolved issues may resurface, necessitating periodic reviews of inactive threads. Lastly, the strong association between toxicity and uncivil features, such as identity attacks and vulgarity, highlights the importance of focusing on these markers for accurate detection, similar to the work done by Ferreira et al.~\cite{ferreira2022how}.






Once toxicity occurs in a conversation, the damage to the participants and the community has already taken place~\cite{xia2020exploring, wulczyn2017ex}. In this study, our focus is derailment, i.e., understanding when a conversation derails and is likely to turn toxic. By detecting derailment, we could potentially avoid toxicity altogether. In the next section, we aim to understand the properties of derailment on GitHub.

\section{Characteristics of Derailment on GitHub}\label{properties_derailed_conversation}

Of the 202 toxic threads, 127 included a preceding uncivil comment (i.e., a derailment point) before the first toxic comment, which we refer to as \DTD. The remaining 75 exhibited sudden toxic comments mid-conversation without any observable derailment point. To analyze derailed conversations, we focus on the 127 threads in {\TCD} with derailment points, referred to as \DTD. In contrast, the remaining 75 threads represent instances of abrupt toxicity.

\subsection{Timing and Distance to Derailment Points} We observe that the median number of comments from the derailment point to the toxic comment is 2. 
The close proximity between derailment and toxic comments suggests that once a thread derails, it is likely to rapidly devolve into toxicity. This aligns with Cheng et al.'s findings, which indicate that negative context and mood increase the likelihood of trolling behavior~\cite{cheng2017anyone}.

The timing of the first toxic comment relative to the derailment point provides additional insights. Considering, 8-hour workday, we observe more than half of the time (58.26\%, 74/127) of toxic comments occur within 8 hours of the derailment comment~\cite{chang2019trouble}, emphasizing the importance of timely intervention. 

\subsection{TBDFs in Derailment Points}
The TBDFs at derailment points ($\geq$10\%) are: Bitter Frustration: 44.88\% (57/127), Impatience: 18.11\% (23/127), and Mocking: 10.23\% (13/127). This and the analysis in Section~\ref{toxic_uncivil_tbdf} indicate that while Bitter Frustration and Impatience are not often toxic themselves, they often serve as precursors to toxicity. Identifying and addressing these TBDFs early on could prevent uncivil exchanges from escalating into full-blown toxicity.

\subsection{Linguistic Features}\label{lang_features}
We analyze language indicative of conversation derailment. Using the 127 derailment point comments, we identified the 200 most frequent unigrams (i.e., words), excluding articles, particles, and common prepositions. Two authors collaboratively categorized (see Table~\ref{tab:discourses}) the remaining 104 unigrams into linguistic features using the card sorting method~\cite{schreier2012qualitative}, where they met in person, discussed, and resolved differences, consulting a dictionary as needed. Based on these categorizations, we automatically counted the frequency of each unigram in the derailment point comments after applying basic preprocessing steps (e.g., tokenize and lemmatize).

Of the 127 derailment point comments, 66.14\% (84/127) used second person pronouns~\cite{levy2022understanding} and 63.78\% (81/127) used first person pronouns. Additionally, 50.34\% (64/127) used both pronouns. Interestingly, these percentages are slightly lower than those found in toxic comments and higher than in the representative GitHub comments, as noted in~\ref{toxic_addressing}. We also found that in derailment points the use of negation terms (`not', `no', etc.), ``WH" questions (`what', `why', `how', `where', etc.), reasoning terms (`because', `since', etc.), communication verbs (`say', `comment', `tell', etc.), and emphasis terms (`actually', `really', etc.) are comparatively higher than in general comments. These elements also occur more frequently in comments marked with incivility TBDFs. 
Table~\ref{tab:discourses} shows the percentages in derailment points along with TBDFs.

\begin{table}[tb]
    \caption{Lexical cues in derailment point comments.}
    \centering
    \begin{tabular}{|p{2.9cm}|p{0.8cm}|p{1.5cm}|p{0.8cm}|}
        \hline
         Linguistic & All & Derailment  & TBDF   \\ 
         features & (8873) & point (127)  & (1025) \\ \thickhline
         Second person & 40.83\%  & 66.14\% & 66.15\% \\ \hline
         WH Question & 40.24\%  & 55.18\% & 55.32\% \\ \hline
         Negation terms & 39.79\% & 62.20\% & 58.44\% \\ \hline
         Reasoning terms & 27.16\% & 41.73\% & 39.80\% \\ \hline
         Emphasis terms & 26.73\% & 40.94\% & 42.54\% \\ \hline
         Communication verbs & 21.89\% & 34.65\% & 36.78\% \\ \hline
    \end{tabular}
    \label{tab:discourses}
\end{table}

\begin{observecomment}
Derailment points in GitHub conversations frequently featured second person pronouns (64.17\%), first person pronouns (65.83\%), or both first and second person pronouns (51.67\%). The use of negation terms, `WH' questions, reasoning terms, communication verbs, and emphasis terms was also notably higher in these derailment points than in the {\NTD}.
\end{observecomment}

\subsection{Trigger Types} Eshani et al.~\cite{ehsani2024incivility} annotated incivility triggers, i.e., what initiated the incivility in the conversation. Following their methodology, two authors of this paper independently annotated triggers at derailment points, achieving a Cohen's Kappa score of 0.78. Disagreements were resolved through in-person discussions for complete agreement.

The most prevalent trigger was `Failed Use of Tool/Code or Error Messages' followed at 30.71\% (39/127), where tool difficulties or bug troubleshooting led to derailment. For example: \textit{``[CODE SNIPPET] ... What more proof do you need? That is everything."} Frustrated tones in this comment about replicating an error led to toxic comments. 
`Technical Disagreement' made up 25.20\% (32/127) of cases, involving disputes over project changes. For instance, \textit{``[CODE SNIPPET] Ask yourself what **intention** it expresses. This is some kind of esoteric gibberish without reference to the subject area. The code is too low-level and [...]"}. In this case, disagreements about method name changes escalated to toxicity. 
`Communication Breakdown' accounted for 22.83\% (29/127) of cases. This included misunderstandings, misinterpretations, typos, or language barriers causing perceived hostility. For example, \textit{``It is impolite to assume that each user opening an issue is stupid and lazy. Of course, I search the issue tracker. I assume I used the wrong keywords. [...]"} Here, a misunderstanding between the commenters triggered incivility, which resulted in toxicity. 
Finally, `Politics/Ideology' accounted for 9.45\% (12/127) of cases, with off-topic political or ideological debates causing derailment. For example: \textit{``Good intentions, but I doubt there's any relation of the origin of the terms blacklist/whitelist to race. There are many idioms and phrases in the English language that make use of colors without any racial backstories. [...]"}.

\begin{observecomment}
The primary triggers at derailment points were `Failed Use of Tool/ Code or Error Messages' (30.71\%), `Technical Disagreement' (25.20\%),  `Communication Breakdown' (22.83\%), and `Politics/Ideology' (9.45\%).
\end{observecomment}
\section{Conversation Derailment Prediction}

Automatically detecting conversational derailment on GitHub is essential for managing the scale and frequency of OSS communication channels. This section describes our method of predicting conversational derailment on GitHub using LLMs. Hua et al. demonstrated that automated derailment forecasting systems perform best when they make their predictions based on {\em Summaries of Conversation Dynamics} (SCD) they have previously generated~\cite{hua2024did}.
SCDs provide a succinct understanding of a conversation's trajectory, detailing the types of interactions that led to its current state and predicting their likely development.
Hua et al. developed a few-shot procedural prompt for SCD generation where the LLM was provided with manually written SCD examples to increase accuracy.
Drawing inspiration from their methodology, we integrate our findings from Section~\ref{properties_derailed_conversation} into the design of SCD generation prompts. We start with Hua et al.'s SCD prompt, customize it for GitHub conversations, and then develop new prompts based on the conversation characteristics observed on GitHub.

\subsection{Baseline Models}
We compare our SCD-based technique to two baselines: 1) CRAFT and 2) Hua et al.'s approach. CRAFT is one of the earliest and best-known models for predicting conversational derailment~\cite{chang2019trouble}. Since its inception in 2019, various other strategies for predicting conversational derailment have been explored~\cite{altarawneh2023conversation,  li2022multimodal, yuan2023conversation, chang2019trouble, kementchedjhieva2021dynamic, hua2024did}. Despite this, the CRAFT model is still a widely used baseline. 
Since our approach adapts Hua et al.'s recent SCD-based technique for predicting derailment on GitHub~\cite{hua2024did}, we also compare to their technique as a baseline.

\subsection{LLM Prompt Design}

We explore several strategies to design the LLM prompt to predict conversational derailment on GitHub. 
Firstly, we adapted Hua et al.~\cite{hua2024did} few-shot procedural SCD prompt for GitHub by specifically mentioning `GitHub conversation' and by providing examples (i.e., few-shot prompting) based on GitHub conversations. An example of the SCD summary we provided in the prompt is as follows:

\smallskip
\begin{quote}
{\em
The conversation involves six users discussing an issue that was encountered in their code. User1 posts the issue asking for guidance, causing User2 to ask for a clarifying detail. User2 provides a solution and User1 responds saying that the solution did not work. User3 then joins the conversation and asks if a solution was ever found. User4 seconds this, causing User5 to join the conversation and comment on how User3 and User4 are not project contributors and therefore have done nothing to try to fix the problem. 
User4 then responds expressing frustration that a solution has not been posted in the last few years.
}
\end{quote}
\smallskip 

While this few-shot prompt appeared to be more effective than Hua et al.'s original prompt at generating SCD summaries for GitHub conversations, we hypothesize that it may not yet be optimal. Hua et al. developed SCD prompts targeting general-purpose conversations, which may not be most effective for the highly technical discussions found on GitHub. We explored whether decomposing the problem~\cite{khot2022decomposed} and
integrating the properties of GitHub derailed conversations, uncovered in Section~\ref{properties_derailed_conversation}, could yield better SCDs for predicting derailment on GitHub. Previous research shows that decomposing the prompts into incremental steps enhances the LLM's accuracy~\cite{khot2022decomposed, zhou2022least}.
Therefore, we devise a prompt that is based on the {\em Least-to-Most} (LtM) prompting strategy~\cite{zhou2022least}, which allows us to integrate our insights from Section~\ref{properties_derailed_conversation}. 

For the language features (e.g., questioning, rhetoric), we used the categories defined by Hua et. al.~\cite{hua2024did} (e.g., rhetorical questions, hedging, questioning logic, etc), which is consistent with our finding as described in section~\ref{lang_features}.
Additionally, to better capture the tone of the conversation, we incorporate social orientation tags (e.g., Assured-Dominant, Gregarious-Extraverted, Warm-Agreeable, Unassuming-Ingenuous, Unassured-Submissive, Aloof-Introverted, Cold, and Arrogant-Calculating), developed using circumplex theory~\cite{strus2017towards}. These tags capture the levels of power and benevolence expressed by each comment and analyzing their interaction provides insight into the conversation's dynamics. Circumplex theory suggests that social interactions can be described by two dimensions: power and benevolence. Power reflects the extent to which an individual seeks to control, lead, or assert themselves in relationships, while benevolence captures the warmth, friendliness, and positivity of interactions. A recent study by Morrill et al. showed that GPT-4 generated social orientation tags are effective for predicting conversation derailment~\cite{morrill2024social}.  

In our final prompt, we integrate triggers, social orientation tags, TBDFs, and linguistic features in a step-by-step manner by decomposing the problem into smaller parts. We used the social orientation tags definitions provided by Morrill et al.~\cite{morrill2024social}, TBDFs and trigger definitions as provided by Ehsani et al.~\cite{ehsani2024incivility}.
Finally, we used the following least-to-most prompt for SCD generation:

\begin{promptbox}{Least-to-most SCD Generator Prompt}
{
\small
\noindent
Here is step-by-step guideline to write an GitHub conversation trajectory summary:

\textbf{Step 1: Identify the main elements of the conversation.}

\textbf{Step 2: Find any triggers of tension in the conversation. The common triggers are:}
\begin{itemize}
    \item \textbf{Failed use of tool/code or error messages:} trouble with code/tool.
    \item \textbf{Communication breakdown:} being misinterpreted by people or being unable to follow.
    \item \textbf{Politics/ideology:} arising over politics or ideology differences (specific beliefs).
    \item \textbf{Technical disagreement:} having differing views on some technical component of the project.
    \item [...]

\end{itemize}

\textbf{Step 3: If there are triggers, identify the social orientation.}
Social orientation from circumplex theory is a social theory that characterizes interactions between speakers. [...] Definitions are:
\begin{itemize}
    \item \textbf{Assured-Dominant:} Demands attention, is firm, self-confident, assertive, persistent, and not self-conscious.
    \item \textbf{Warm-Agreeable:} Interested in people, polite, cooperative, accommodating, gentle.
    \item \textbf{Arrogant-Calculating:} Boastful, manipulative, cunning, cocky.
    \item [...]


\end{itemize}

\textbf{Step 4: Describe the sentiments and tones expressed by each participant. Indicators include:}
\begin{itemize}
    \item \textbf{Bitter frustration:} strong frustration.
    \item \textbf{Impatience:} feeling that resolution is taking too long.
    \item \textbf{Insulting:} directed insults.
    \item [...]
\end{itemize}

\textbf{Step 5: Note conversation strategies to find tones.}
\begin{itemize}
    \item \textbf{Rhetorical Questions:} posed for a point, not an answer.
    \item \textbf{Posing Challenges/Clarifications:} asking for elaboration.
    \item \textbf{Hedging:} softening statements.
    \item [...]
\end{itemize}

\textbf{Step 6: Combine information from steps 1-5 to write a short summary.}

\textbf{Step 7: Refine the summary using the following points:}
\begin{itemize}
    \item Do not include specific technical details.
    \item Keep it concise, capturing key moments and tonal shifts.
    \item Focus on moments where conversation dynamics change.
\end{itemize}

\textbf{Example:} "Several contributors discuss an unresolved code issue. User1 requests guidance, [...] responds with frustration over the lack of a solution, and the conversation ends on a tense note."
\newline

\textbf{Task:} Using the guideline, write the final trajectory summary for a provided GitHub conversation transcript.
\newline

Conversation Transcript: $<insert\ conversation>$
\newline

Write only the trajectory summary in double quotes.
}

\end{promptbox}

Here is an example SCD summary generated using this prompt:

\smallskip
\begin{quote}

{\em 
The conversation starts with @USER1 seeking documentation for brewfile syntax, which @USER2 provides partially by pointing to the DSL definition. However, @USER1 expresses frustration over the lack of comprehensive documentation and @USER3 responds defensively, stating that issues should not be left open without actionable work. The tone shifts as @USER1 criticizes the lack of documentation, leading @USER4 to accuse him of violating the Code of Conduct. Despite @USER1's attempts to clarify his intentions, the conversation ends with a tense exchange and @USER1 deciding to leave.
}
\end{quote}

\smallskip

We observe that this SCD summary was more effective at capturing the emotional aspects and tone of the GitHub conversations, compared to our initial approach. We employ a follow-up predictor prompt based on the generated SCD summary asking the model to assess the likelihood of the conversation derailing into toxicity on a scale from 0 to 1. This second prompt is structured as follows:

\begin{promptbox}{Predictor Prompt}

{
\small
You will be provided with a summary of a GitHub conversation trajectory. 

A Conversation Trajectory Summary captures the evolution of a discussion by identifying key elements, triggers of tension, conversational strategies, interaction patterns, sentiments, and tonal shifts among participants. It provides a concise overview of significant dynamics and tone changes without detailing the technical aspects or specific arguments.
\newline

Now based on the given GitHub conversation trajectory summary, what is the probability that the conversation may derail and become toxic on a scale of 0 to 1 (Toxicity is defined by “any rude, disrespectful, or unreasonable comment that is likely to make one leave a discussion.”)?
\newline

Conversation trajectory summary: $<insert\ summary>$
\newline

Write exactly one word: the probability rounded to two decimal places.
\newline

Do not write reasoning.
}

\end{promptbox}

This two-step prompting process allows us to first generate a comprehensive summary of the conversation and then use that summary to make a more informed prediction about the potential for toxicity.

\subsection{Experiment Setup and Metrics}

We conduct all of our experiments using the \textit{LLaMA-3.1-70B} model, as it is one of the best open-source state-of-the-art LLMs at the time of writing. We set the model temperature to 0 to minimize output variance and a context window size of 8192. We use popular metrics to evaluate classification: Precision, Recall and F1-score.

\begin{itemize}
    \item Precision refers to the proportion of true positive observations among all the predicted positive observations. 
    \[Precision = \frac{True\ Positive}{True\ Positive + False\ Positive}\]
    \item Recall represents the proportion of true positive observations out of all actual positive observations in the "true" class.
    \[Recall = \frac{True\ Positive}{True\ Positive + False\ Negative}\]
    \item The F1-score is the harmonic mean of Precision and Recall, providing a balanced measure of both metrics.
    \[F1-score = 2\ *\ \frac{Recall\ *\ Precision}{Recall\ +\ Precision}\]
\end{itemize}

For each toxic conversation in our dataset, we provide all the comments up to, but excluding, the first toxic comment. 

\subsection{Results and Discussion}
We conduct experiments over using the dataset consisting of the 127 derailed toxic threads and 696 non-toxic threads, a total of 823 data points. The results of this experiment are shown in Table~\ref{tab:models_accuracy}. 
As previous research shows that the decision threshold can vary widely for different datasets in conversational derailment prediction~\cite{chang2019trouble}, we include results from different thresholds: $T \geq $ 0.4, 0.5, and 0.6. The CRAFT model consistently underperformed across all thresholds, while the three prompt-based techniques demonstrated superior results. Among these, the GitHub-specific prompts significantly outperformed the generic approaches at all thresholds.

\begin{table}[tb]
    \caption{Derailment prediction results for different models on Derailed Dataset and Non-toxic Dataset. (SCD = Summaries of Conversation Dynamics, $T$ = $Threshold$).}
    \centering
    \begin{tabular}{|l|c|c|c|c|}
        \hline
        \multirow{1}{*}{\textbf{Model}} 
        & \multirow{1}{3em}{\centering $T$ ($\geq$)}
        & \multirow{1}{4em}{\centering{Precision}} 
        & \multirow{1}{3em}{\centering{Recall}}
        & \multirow{1}{3em}{\centering{F1}}\\
        \thickhline
        \multirow{3}{5em}{{CRAFT~\cite{chang2019trouble}}} & 0.4 & 0.20 & 0.76 & 0.32 \\ 
        & 0.5 & 0.27 & 0.58 & 0.37 \\ 
        & 0.6 & 0.33 & 0.47 & 0.39 \\ 
        \hline
        \multirow{3}{5em}{{Hua et al. SCD~\cite{hua2024did}}} 
        & 0.4 & 0.64 &	0.54 &	0.59 \\
        & 0.5 & 0.80 &	0.35 &	0.48 \\
        & 0.6 & 0.85 &	0.32 &	0.47 \\
        \thickhline
        \multirow{3}{5em}{{Few-shot SCD}} 
        & 0.4 & 0.45 &	0.83 & 0.58 \\
        & 0.5 & 0.60 &	0.68 & 0.63 \\
        & 0.6 & 0.60 &	0.66 & 0.62 \\
        \hline
        \multirow{3}{5em}{{Least-to-most SCD}} 
        & 0.4 & 0.58 &	0.81 &	0.68 \\
        & 0.5 & 0.76 &	0.65 &	0.70\\
        & 0.6 & 0.79 &	0.61 &	0.68 \\
        \hline
    \end{tabular}
    \label{tab:models_accuracy}
\end{table}

The Least-to-Most SCD prompt consistently achieved the highest F1-scores at each threshold, with particularly strong performance in precision compared to the Few-shot SCD prompt. This precision advantage is critical, as higher precision ensures fewer false positives, which is particularly valuable in real-world scenarios where non-toxic conversations vastly outnumber toxic ones~\cite{raman2020stress}. We envision a threshold-based intervention strategy to mitigate toxicity: higher thresholds could alert moderators to review flagged content, while lower thresholds could trigger automated bots to issue reminders promoting civil discourse.


\subsection{Error Analysis}~\label{sec:error}

We limited the error analysis to the least-to-most SCD prompt using the threshold of 0.5, which has the highest F1-score.

To better understand the performance and limitations of the model and datasets,  we conduct an error analysis that focuses on two types of errors:
1) 44 cases where the model predicted toxic conversations as non-toxic; and
2) 26 cases where the model predicted non-toxic conversations as derailing to toxic.

Two authors of the paper reviewed the conversations, examined the generated SCD, and determined the most likely reason for the error. They finalized the error categories using card sorting~\cite{schreier2012qualitative}, with some cases belonging to more than one category. 

For instances where the model incorrectly predicted non-toxic conversations as derailing into toxicity, the primary error categories were: \textit{acknowledges tensions but overestimates effect} (9/26 cases), where the model correctly identified tensions in a comment but overestimated their impact, often in cases of technical disagreements; \textit{misinterpretation of tones in comments} (5/26 cases), where the model misinterpreted positive or neutral comments as negative, such as perceiving constructive feedback as dismissive; and \textit{the issue/PR was locked/closed} (5/26 cases), where the last comment in the conversation, often from automated bots or users, announced the locking or closure of the issue or pull request. This caused the model often to predict the conversation is going to be toxic.

In cases where the model failed to predict toxicity in conversations that derailed, the key error categories were: \textit{underestimating or overlooking the seriousness of tone or comment} (20/44 cases), where subtle toxic signals were missed or misinterpreted, such as expressions of frustration; \textit{conversation context being too extensive for effective analysis} (8/44 cases), where the length and complexity of discussions surpassed the model's ability to capture relevant nuances, even with a large context window; and \textit{the derailed uncivil comment is followed by a civil comment} (4/44 cases), where toxic comments appeared after a civil comment. In the later category, it is likely that the juxtaposition of civility and incivility can introduce ambiguity, making it difficult for the model to predict accurately.

\subsection{Implications and Recommendations}

Our study indicates that developing domain-specific approaches to address toxicity in software engineering contexts can offer strong improvements over generic methods. 
The error analysis reveals that the moderators sometimes failed to lock toxic issues and pull requests. This oversight can be mitigated by implementing an automated proactive moderation system, which can identify and flag potentially toxic interactions early, prompting moderators to take timely action.
To operationalize our findings, we propose creating real-time monitoring tools for proactive moderation based on thresholds, such as interactive dashboards and automated bots that can flag at-risk conversations and gently remind users to follow codes of conduct~\cite{qiu2023climate, li2021code}. For example, a threshold-based intervention strategy can be adopted to mitigate toxicity. For instance, at higher thresholds, moderators could be prompted to manually review flagged content requiring immediate attention. Conversely, at lower thresholds, automated bots could issue reminders to encourage civil discourse. The generated SCDs also offer an opportunity for more transparent and explainable moderation practices, providing clear justifications for interventions. 
To address root causes of derailment, we suggest developing targeted interventions based on common triggers and linguistic features associated with toxic interactions. 
Community education efforts should include resources explaining patterns of conversational derailment and strategies for maintaining constructive dialogue. 

\section{Related Work}

Our work builds upon and extends previous research in two main areas: toxicity analysis in software engineering and conversational derailment prediction.

\subsection{Toxicity analysis in SE artifacts}

Researchers have investigated negative interactions, such as offensive language, sentiments, emotions, incivility, tones, and toxicity, and developed automated tools to detect them across different OSS channels, including pull requests, issues, code reviews, Stack Overflow, chat forums and GitHub discussions~\cite{sarker2020benchmark, sarker2023automated, sarker2023toxispanse, sarker2022identification, ebert2019confusion, novielli2020can, cheriyan2021towards, ferreira2021shut, ferreira2022how, ferreira2024incivility, islam2018sentistrength, calefato2018sentiment, novielli2014towards, ortu2015bullies, chen2019sentimoji, cohen2021contextualizing, imran2022data, imran2024shedding, imran2024uncovering, serebrenik2022social, qiu2022detecting,  gunawardena2022destructive, hata2022github, almarimi2023improving, ehsani2023exploring, ehsani2024incivility, madampe2023framework, wang2023multimodal, coutinho2024looks, tian2024analyzing, rahman2024words}. 
Early software engineering research predominantly focused on sentiment and emotion analysis within software engineering communities
~\cite{garcia2013role, novielli2014towards, ortu2015bullies, gachechiladze2017anger, islam2018sentistrength, calefato2018sentiment, chen2019sentimoji, egelman2020predicting}. 
In recent years, researchers' focus has shifted towards incivility and toxicity~\cite{prana2021including, cohen2021contextualizing, sarker2023automated, ehsani2024incivility, ferreira2024incivility}.

Raman et al. examined OSS contributors' stress and burnout, linking negative interactions to increased dropout rates among developers~\cite{raman2020stress}. 
Sarker et al. conducted experiment on automatically detecting toxicity in code reviews~\cite{sarker2023automated, sarker2022identification, sarker2023toxispanse}.
Jamieson et al. studied the role of value-related interactions in contributor turnover~\cite{jamieson2024predicting}. 
Miller et al. explored the dynamics of toxic interactions in OSS projects and emphasized the importance of active moderation and the potential benefits of automated tools for early detection of toxic behavior~\cite{miller2022did}. 
Heish et al. examined various moderation strategies and assessed how bots can be helpful~\cite{hsieh2023nip}.
Ferreira et al. and Ehsani et al.~\cite{ferreira2022how, ehsani2024incivility} analyzed locked GitHub issues to understand the causes and patterns of incivility and provided a comprehensive dataset of uncivil conversations.

\subsection{Conversation derailment}

Online conversations often derail into toxicity, leading to negative user experiences and increased moderation challenges. In recent years, studies have explored methods for predicting and mitigating derailment and toxicity~\cite{zhang2018conversations, levy2022understanding, de2021beg, bao2021conversations, chang2019trouble,  xia2020exploring, kementchedjhieva2021dynamic, li2022multimodal, yuan2023conversation, altarawneh2023conversation, hua2024did}.

Chang et al.~\cite{chang2019trouble} developed the CRAFT tool to detect conversational derailment by analyzing the flow of past comments. They tested the effectiveness of the tool on Wikipedia and Reddit datasets.
Kementchedjhieva et al.~\cite{kementchedjhieva2021dynamic} used a BERT-based model on the same datasets, while Leung et. al.~\cite{leung2023hashing} predicted whether Twitter conversations would become unhealthy. 
ConvoWizard provided users on Reddit's r/ChangeMyView with toxicity forecasts, which most users found very useful~\cite{chang2022thread}.
Multimodal approaches have also been explored~\cite{li2022multimodal}.
A study examining the structure of toxic conversations on Twitter revealed that toxicity begets toxicity, and that toxic exchanges have larger reply tries but sparse follow graphs~\cite{saveski2021structure}. 
Considering conversation context reduces false positives and negatives in toxicity prediction~\cite{anuchitanukul2022revisiting}. 
Xia et al. identified user propensity, previous toxicity, engagement volume, and community norms as key antecedents of toxicity~\cite{xia2020exploring}. 
In another case, early controversy and conversational deterioration forecasts have been enabled by features from the start of a conversation~\cite{hessel2019something}, `edit' tokens in comments~\cite{de2021beg}, and patterns of consecutive user behaviors~\cite{tshimula2020predicting}.

Despite advancements, preemptive toxicity detection datasets and models still show limitations~\cite{schluger2022proactive}. 
Recent work has explored hierarchical transformers with multitask learning~\cite{yuan2023conversation}, 
GNN~\cite{altarawneh2023conversation}, social orientation features~\cite{morrill2024social}, 
and conversation summarization using the latest GPT models~\cite{hua2024did}.


\section{Threats to Validity}

We note potential threats to the validity of our study in the following categories: construct validity, internal validity, and external validity.

\subsection{Construct validity} This concerns whether our study accurately measures its intended concepts. Toxicity and conversational derailment are subjective, and open to varied interpretations. Our use of \textit{GPT-4o} for annotation, and \textit{LLaMA-3.1:70B} for summarizing and predicting toxicity may introduce biases from its training data. We performed rigorous checking during annotation and conducted error analyses to identify these biases.

\subsection{Internal validity} This relates to the accuracy of our findings, free from external influences. Biases and inconsistencies in our annotation process are potential threats. Despite using multiple annotators and cross-checking, human error and subjective judgment may affect our results. We address this in various ways: through rigorous guidelines for annotators, achieving high inter-annotator agreement, and utilizing a model-in-the-loop approach. 

\subsection{External validity} This addresses the generalizability of our findings. Using only GitHub data may limit applicability to other OSS platforms. Thus, our findings may not directly transfer to platforms like JIRA or non-OSS forums. However, our prompt design methodology is adaptable to any domain. Including diverse datasets from various platforms in future research will help validate and extend our conclusions.

Another threat the external validity is the limited dataset that we curated. We note that our empirical observations of toxicity and derailment on GitHub need to be further investigated on a larger scale as well as in specific GitHub sub-communities, as our findings may be limited to this specific curated dataset.
\color{black}

\section{Conclusion}
In this study, we aim to understand and predict conversational derailment and toxicity on GitHub. Our analysis reveals that users are more likely to initiate toxic conversations, while higher developer engagement correlates with lower toxicity levels. We also find that toxic conversations tend to be longer and escalate quickly after derailment. Generating conversation trajectory summaries using LLMs, we propose a proactive moderation approach, achieving F1-score of 0.70 with 0.76 precision in predicting conversational derailment.

Future work should focus on enhancing our understanding of conversational derailment and toxicity on GitHub and similar platforms. Expanding the dataset to include conversations from additional OSS platforms like JIRA, Gitter and other discussion boards will help validate our findings across different environments and improve their generalizability. Developing and deploying real-time intervention tools that provide immediate feedback during conversations can prevent the escalation of toxic interactions. Understanding the most relevant social orientation tags in the OSS context can be another line of future work. 

\bibliographystyle{IEEEtran}
\bibliography{references}

\end{document}